\documentclass[a4paper]{article}

\usepackage{subfiles}
\usepackage{graphicx}
\usepackage{color}
\usepackage{algorithm}
\usepackage{algorithmic}
\usepackage{amsmath}
\usepackage{amsfonts} 
\usepackage{amsthm}
\usepackage{cases}
\usepackage[colorlinks=true]{hyperref}
\usepackage{caption}
\usepackage{subcaption}
\usepackage{color}
\usepackage{longtable}
\usepackage{subfiles}
\usepackage{typearea}

\typearea{12}

\usepackage{tcolorbox}

\theoremstyle{definition}
\newtheorem{problem}{Problem}

\newtheorem{remark}{Remark}
\newtheorem{theorem}{Theorem}
\newtheorem{example}{Example}

\newcommand\reff[1]{Fig.~\ref{#1}}
\newcommand\reft[1]{Table~\ref{#1}}
\newcommand\refa[1]{Algorithm~\ref{#1}}
\newcommand\refs[1]{Section~\ref{#1}}
\newcommand\refss[1]{Section~\ref{#1}}
\newcommand\refappendix[1]{Appendix~\ref{#1}}

\newcommand\Cell[1]{\begin{tabular}{l}#1\end{tabular}}

\newcommand\MAGENTA[1]{\textcolor{magenta}{#1}}
\newcommand\LineBreak{\MAGENTA{/}}

\newcommand\Token[1]{t_\texttt{#1}}
\newcommand\Text[1]{\text{``}\texttt{#1}\text{''}}

\newcommand\bbR{\mathbb{R}}
\newcommand\Range[2]{\{#1,\ldots,#2\}}
\newcommand\InRange[2]{\in\Range{#1}{#2}}

\newcommand\unom{u_\mathrm{nom}}

\newcommand\calT{\mathcal{T}}

\newcommand\calX{\mathcal{X}}
\newcommand\calS{\mathcal{S}}
\newcommand\barcalS{\bar{\mathcal{S}}}
\newcommand\Concat{\texttt{concat}}
\newcommand\ktop{k_\mathrm{top}}

\newcommand\FCBF{F_\mathrm{CBF}}

\newcommand\FNC{F_\mathrm{NC}}

\title{CBF-LLM: Safe Control for LLM Alignment}
\author{Yuya Miyaoka, Masaki Inoue}
\date{}

\begin{document}

\maketitle


\begin{abstract}
This paper proposes a control-based framework for aligning large language models (LLMs) by leveraging a control barrier function (CBF) to ensure user-desirable text generation. 
The presented framework applies the safety filter, designed based on the CBF, to the output generation of the baseline LLM, i.e., the sequence of the token, with the aim of intervening in the generated text.
The overall text-generation system is implemented with Llama~3 and a RoBERTa model, and the source code is available at \url{https://github.com/Mya-Mya/CBF-LLM}. 
The experiment demonstrates its control ability and effectiveness in reducing the number of interventions needed for user-specified alignment tasks.
\end{abstract}


\section{Introduction}\label{S:Introduction}
While large language models (LLMs) are known to have strong language understanding and generation abilities,
they can also generate harmful, biased, and toxic content \cite{Shen23}\cite{Minaee24}.
Alignment of LLMs ensures that they generate content that is ``desirable'' for the user, typically meaning content that is safe and ethical.
Various approaches for LLM alignment have been presented (\cite{Shen23}, \cite{Minaee24}, \cite{Wang23} and reference therein).

The major approach to the alignment is reinforcement learning from human feedback (RLHF) \cite{Ouyang22}, 
where a reward model is constructed by human feedback and used for the training of LLMs.
Variants of RLHF architectures are also proposed, such as Safe-RLHF \cite{Dai24}, SENSEI \cite{Ruibo22}, and f-DPG \cite{Go23}, 
and their implementations are presented, such as training pre-trained LLMs \cite{Bai22}\cite{Chunting23}, and applications like information-seeking chatbot \cite{Glaese22}.
Collecting human feedback with data is time-consuming and expensive.
To overcome the drawback, RL from AI Feedback (RLAIF) is presented in \cite{Bai22b} instead of using human labels. In addition, the method to construct the training data automatically is proposed in \cite{Kim23}.
Furthermore, to reduce the computational cost, direct preference optimization (DPO) is proposed \cite{Rafael23}, where the training data is directly used for training LLMs without accessing the reward model.
Supervised fine-tuning (SFT) is a different approach for alignment from RLHF, as studied in \cite{Liu23}. 
A common feature of alignment methods like RLHF and SFT is that they modify LLMs' model parameters.

An alternative approach for LLM alignment is to directly intervene in the input prompt or output of LLMs, rather than modifying the model parameters.
In-context learning (ICL) \cite{Dong24} is a major approach for intervening in the input prompt. 
In ICL, a few demonstrations are provided in prompt to instruct the LLMs on the task, including few-shot learning \cite{Tom20}\cite{Zhao24}.
As the methods for intervening in the output,
the work \cite{Cao21} proposes a method to format output for retrieval application, and the work \cite{Keskar19} proposes a repetition penalty to prevent LLMs from repeating the same words and expressions.
In addition, Transformers module provides some functions to modify the output, such as \texttt{NoBadWordsLogitsProcessor} and \texttt{MinLengthLogitsProcessor} \cite{HuggingFaceND}.

\begin{figure}[h]
    \centering
    \includegraphics[width=\linewidth]{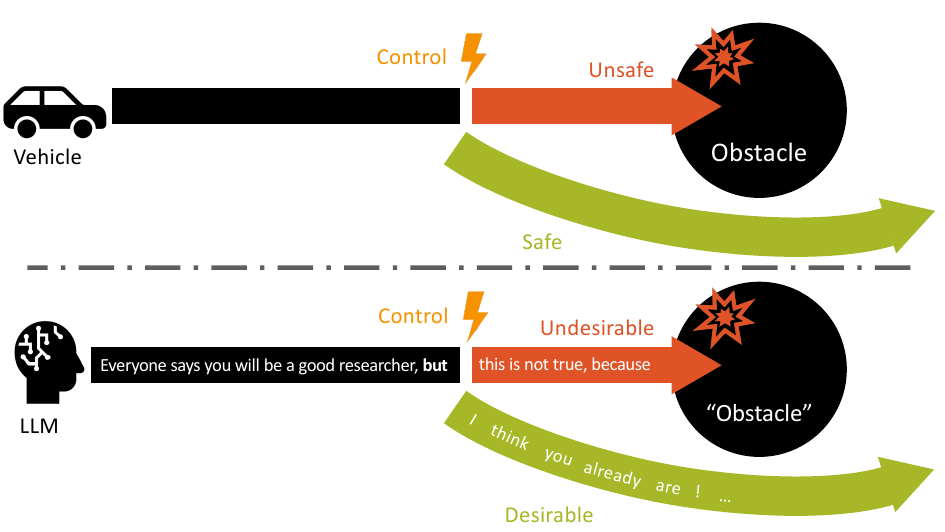}
    \caption{Concept of CBF-LLM. 
    Upper: Collision avoidance in a vehicle control system,
    Lower: \textit{Collision avoidance} in text-generation by LLMs.
    }
    \label{F:Introduction.Concept}
\end{figure}

One can view that the intervention approach to alignment, which avoids undesirable output, as analogous to ``collision avoidance'', the most fundamental problem in control engineering. 
In control engineering, various studies are conducted for safety assurance, including collision avoidance \cite{Axton24}\cite{Dawson23}\cite{Nishimura24}.
A promising approach for collision avoidance is the control barrier function (CBF), as studied in theoretical works \cite{Ames19}\cite{J20}\cite{Brett21} and in real-world applications \cite{Juqi23}.
An analogy between vehicle collision avoidance and intervention-based LLM alignment can be drawn as illustrated in \reff{F:Introduction.Concept}.
The goal of vehicle collision avoidance is to prevent collisions with obstacles by intervening in the vehicle's trajectory.
For example, if there are obstacles ahead of the vehicle, it is necessary to operate the steering or use the brakes to avoid colliding with them.
Similarly, the goal of LLM alignment is to prevent undesirable outputs, such as harmful or biased content, by intervening in LLM's trajectory, i.e., the sequence of generated tokens.
Consider the LLM as an analogy to a vehicle, and the generated text as an analogy to the vehicle's trajectory.
Both vehicle collision avoidance and LLM alignment aim to guide the complex system away from undesirable states by designing proper control strategies.
Based on this analogy, this paper develops a control strategy for LLM alignment based on the CBF,
focusing on intervening in the trajectory of generated text, i.e., the sequence of generated tokens from LLMs.

This paper proposes a framework for control-based LLM alignment by applying a safety filter that intervenes in the LLM output to generate the user-desirable outcomes.
To this end, we leverage the idea of the CBF to improve the safety and controllability of the outputs of LLMs.
We aim to design the CBF-based safety filter that intervenes in the output of LLMs into the user's desired content.
The CBF filter and the baseline LLM constitute a novel text generation system, which we call ``CBF-LLM''.
This paper also conducts the text-generation experiment on CBF-LLM. 
In the experiment, the CBF control enabled the alignment task to be completed with fewer interventions. 

The contributions of this paper are as follows:
\begin{itemize}
    \item The proposed CBF-LLM is realized in an add-on manner to a baseline LLM: an external filter is simply added to the LLMs 
    without accessing their model parameters.
    In this sense, CBF-LLM is a ``learning-free'' framework for aligning LLMs.
    CBF-LLM is broadly applicable to various LLMs, as is designed independently of the underlying LLM.
    \item The concept of CBF-LLM is implemented with Llama~3 and a RoBERTa model in \refs{S:E}. The source code can be accessed in \url{https://github.com/Mya-Mya/CBF-LLM}.
\end{itemize}
This paper also attempts to bridge the gap between control engineering and NLP. 
While the \textit{theoretical analysis} of LLMs, such as reachability, has been studied in the works \cite{Aman24} and \cite{Soatto23},
this paper presents a \textit{design} method of LLM-based text-generation systems.

The rest of this paper is organized as follows.
In \refs{S:PL}, the basic theory of CBF is provided, and the structure of nominal LLM is reviewed.
In \refs{S:CBFLLM}, the concept and design of CBF-LLM are proposed.
In \refs{S:E}, the experiment of CBF-LLM is conducted.
Finally, in \refs{S:Conclusion}, the conclusion of this paper is presented.

Notation: symbol $V[i]$ represents the $i$-th element of the vector $V$.


\section{Preliminary}\label{S:PL}

\subsection{Control Barrier Function for Safe Control}\label{SS:CBF}
Control barrier function (CBF), developed in the control community, provides safety assurance in control systems \cite{Ames19}. 
This section briefly reviews CBF and CBF-based safety control.

Consider the following dynamical system to be controlled:
\begin{align}
    \dot x = g(x,u) ,
    \label{E:PL.NominalDynamics}
\end{align}
where $x\in\bbR^n$ is the state variable of the object being controlled, and $u\in\bbR^m$ is the action applied to the object.
The function $g$ represents the system dynamics; how this object is affected by the current state $x$ and action $u$.
A typical example of the system \eqref{E:PL.NominalDynamics} includes a vehicle dynamics,
where the state $x$ is coordinate, velocity, angle, etc, and the action $u$ is accelerator pedal depression, steering angle, etc.

We aim to design the assisted control system with safety assurance.
As for safety, we let the safe and unsafe sets be denoted by $\calS\subseteq\bbR^n, \barcalS\subseteq\bbR^n$, respectively.
Then, the safety means to constrain $x$ within the safe set $\calS$, i.e., $x\in\calS$.
Consider that the nominal action $\unom\in\bbR^m$ is provided which might violate the safety, i.e., $\unom$ might generate the unsafe state $x\in\barcalS$.
Then, we address the problem of designing ``safety filter'' $F:\bbR^m\to\bbR^m$, as follows:
\begin{problem}[Safety Filter]
    Find the safety filter $F:\bbR^m\to\bbR^m$ such that the system \eqref{E:PL.NominalDynamics} with $u=F(\unom)$ generates $x(t)\in\calS$ for all nominal actions $\unom$ for all time $t$.
\end{problem}
As a preliminary, we design a continuously differentiable function $h:\bbR^n\to\bbR$, called a ``constraint function'', such that
\begin{align}
    h(x) \ge 0 , \quad & x \in\calS, \label{E:PL.SafeSet}\\
    h(x) < 0, \quad & x \in\barcalS \label{E:PL.DangerSet}
\end{align}
hold.
The safety is equivalent to the constraint: $h(x) \ge 0$.
In the assisted driving example, the nominal action $\unom$ is a manual action by the driver, and the unsafe set $\barcalS$ includes the positions of obstacles like pedestrians. 
The function $h$ can be the distance between the vehicle and the obstacle. 
With the manual action by the driver $\unom$, the vehicle may enter the unsafe set, such as colliding with obstacles.
The safety filter $F$ needs to modify $\unom$ to output $u$ such that $h(x)\ge0$ holds, i.e., the trajectory in which the vehicle never collides with obstacles.

To construct the safety filter $F$ that keeps the safety constraint $h(x)\ge0$,
the control barrier function filter (CBF filter, \cite{Gurriet18}\cite{Ames19}) is presented. 
The CBF filter intervenes in the nominal action value $\unom$ to introduce a safe state of the object as follows:
\begin{align}
    u = &~ \arg \min_u (\unom-u)^2 , \label{E:PL.CBFObjective}\\
    \text{s.t.} &~ \dot h(x) \ge - \alpha(h(x))  ,\label{E:PL.CBFConstraint}
\end{align}
where $\alpha:\bbR\to\bbR$ is a class-$\mathcal{K}$ function which holds $h(0)=0$ and monotonically increasing, i.e., $\frac{dh(x)}{dx}>0$.
The function $h$ is called a \textit{control barrier function} if there exists $u$ such that the constraint \eqref{E:PL.CBFConstraint} holds.
In addition, \eqref{E:PL.CBFConstraint} is called a CBF constraint.
In the assisted driving example, when the manual action $\unom$ is expected to cause a collision with an obstacle, the CBF filter intervenes in $\unom$ to provide safe driving. 
To constrain the action $u$ by the CBF filter, the following statement holds:
\begin{theorem}
    The state $x$ of the system \eqref{E:PL.NominalDynamics} is in the safe set, i.e., $x\in\calS$ for all time if $h$ is a control barrier function and the action $u$ satisfies the CBF constraint \eqref{E:PL.CBFConstraint}.
\end{theorem}
\begin{remark}
    The objective function \eqref{E:PL.CBFObjective} ensures that the filtered action $u$ remains as close as possible to the nominal action $\unom$.
    In this sense, the CBF filter archives safety by the ``minimum'' intervention.
\end{remark}

The CBF filter is capable of applying in discrete-time systems by re-formulating the CBF constraint \eqref{E:PL.CBFConstraint} as follows \cite{Zeng21}:
\begin{align}
    \Delta h(k) = h(k+1) - h(k) \ge - \alpha (h(k)) ,
    \label{E:PL.DiscreteTimeCBFConstraint}
\end{align}
where $k$ is a discrete time.


\newcommand\fLLM{f_\mathrm{LLM}}

\subsection{Text generation by Large Language Models}\label{SS:LLM}
This section reviews and analyses the text generation by large language models (LLMs) while particularly focusing on their structure.
In this paper, ``text'' means the sequence of tokens and $\calX$ denotes the set of the texts.
The text corresponding to a specific expression in natural language is displayed as $x = x(\Text{<text>})$.
For example, the text $x\in\calX$ for ``Have a nice day.'' is displayed as $x(\Text{Have a nice day.})$.
Each token $t$ is identified by a positive integer, i.e. $t \in\Range{1}{N} =: \calT$, and $N$ is the number of tokens the LLM has. 
The token corresponding to a specific expression in natural language is displayed as a numerical constant $\Token{<token>}$. 
For example, the token for ``dog'' is displayed as $\Token{dog}$.
The function $\Concat:\calX\times\calT\to\calX$ concatenates text and token to output a text. For example,
$
    \Concat(x(\Text{Have a nice}), \Token{day}) = x(\Text{Have a nice day})
$.

\begin{figure}[h]
    \centering
    \includegraphics[width=0.75\linewidth]{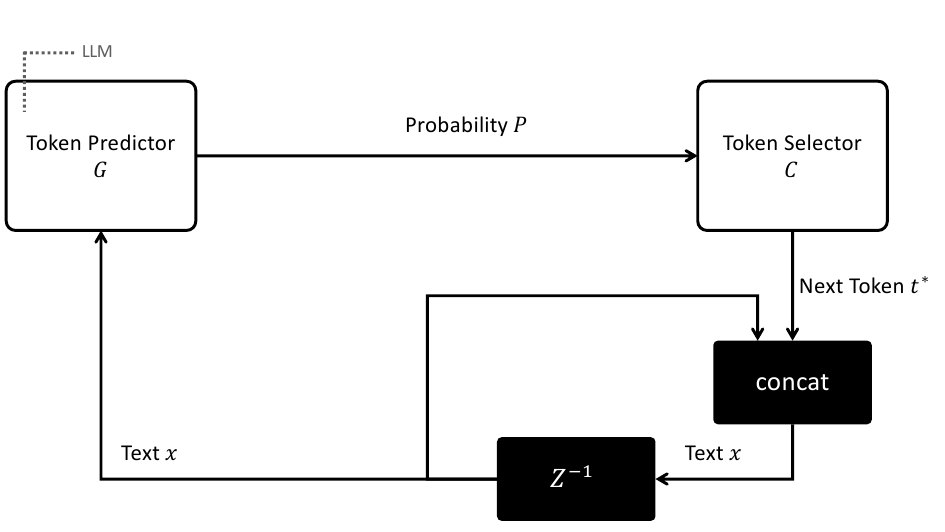}
    \caption{Nominal structure for text generation}
    \label{F:PL.NominalLLM}
\end{figure}

Text generation is performed by iteratively adding a new token, starting from the initial text. 
The text generation by LLM is considered to be a discrete-time dynamical system as shown in the block diagram \reff{F:PL.NominalLLM}.
In this figure, the token predictor $G:\calX\to\bbR^N$ plays a central role in the text generation: 
it drives the input text $x$ to output the probability $P\in\bbR^N$, where $P[t],t\in\calT$ displays how probable that the token $t\in\calT$ would follow by the text $x$. 
This $G$ is typically replaced by LLMs.
The token selector $C$ selects the next token $t^*$ which follows $x$ based on the probability $P$.
Finally, in the concatenator $\Concat$, the new token $t^*$ is concatenated with $x$ to derive the new text $x$.
The $Z^{-1}$ block represents the time delay, playing a role of a memory of text $x$.

Let $x_0\in\calX$ be the initial text, and $k\in\{0,1,\ldots\}$ be the ``time'', which counts the number of tokens added during the generation. Then, the text generation is expressed as follows:
\begin{align}\begin{cases}
    x(0) = x_0, \\
    P(k) = G(x(k)), \\
    t^*(k) = C(P(k)), \\
    x(k+1) = \Concat(x(k), t^*(k)) .
    \label{E:PL.NominalLLMDynamics}
\end{cases}\end{align}
As described above, the text-generation LLM such as GPT-2, and Llama 3 is included in the token predictor $G$. The token predictor not only accommodates the LLM but also has the function of processing the output of the LLM.
Let the text-generation LLM as $\fLLM:\calX\to\bbR^N$. Then, the probability $P$ is calculated as follows:
\begin{align}
    P[i] = \mathrm{softmax}(\fLLM(x)[i]/T) =
    \frac{
        \exp(\fLLM(x)[i]/T)
    }{
        \sum_{j=1}^N \exp(\fLLM(x)[j]/T)
    }
    ,
    \label{E:PL.Probabilification}
\end{align}
where $T\ge0$ is the temperature.

The algorithm of the text generation is given as follows:
\begin{algorithm}[h]
\caption{Nominal text generation}
\label{A:PL.TextGeneration}
\begin{algorithmic}[1]
    \REQUIRE $x_0 \in\calX$ : initial text.
    \REQUIRE $T \ge0$ : temperature, hyperparametes of the token predictor $G$.
    \STATE $k \gets 0$
    \STATE $x(0) \gets x_0$
    \WHILE{\TRUE}
        \STATE $P \gets \mathrm{softmax}(\fLLM(x)/T)$
        \STATE $t^* \gets$ randomly choose the token according to the $P$.
        \STATE $x(k+1) \gets \Concat(x(k), t^*)$
        \STATE $k \gets k+1$
    \ENDWHILE
\end{algorithmic}
\end{algorithm}



\section{CBF-LLM}\label{S:CBFLLM}
This section presents the control-based alignment of text-generation systems and their detailed implementation.
Symbols and their meanings used in this paper are summarized in \reft{T:CBFLLM.NotationMeanings}.
\begin{table}[h]
    \centering
    \caption{Symbols and their meanings}
    \label{T:CBFLLM.NotationMeanings}
    \begin{tabular}{c|l|l}
    \hline
        Symbols & Meaning in CBF & Meaning in LLM Alignment (CBF-LLM) \\
    \hline
        $x$ & 
            \Cell{State of the controlled object, $x\in\bbR^n$.} & 
            \Cell{Generated text, $x\in\calX$.} \\
        $h$ & 
            \Cell{Constraint function, $h:\bbR^n\to\bbR$.} & 
            \Cell{Language-constraint function (L-CF),\\ $h:\calX\to\bbR$.} \\
        $\calS$ & 
            \Cell{Safe state set.} & 
            \Cell{Desirable text set.} \\
        $\barcalS$ & 
            \Cell{Unsafe state set.} & 
            \Cell{Undesirable text set.} \\
    \hline
    \end{tabular}
\end{table}

The alignment discussed in this paper aims to ensure desirable text generation by weak intervention to the output of LLMs.
To clarify the meaning of ``desirable'', we let the desirable and undesirable text sets be $\calS\subseteq\calX$ and $\barcalS\subseteq\calX$, respectively, based on the respective alignment goals.
\begin{example}
Suppose that the alignment goal is set to generate non-toxic content. 
Then, $\calS$ is the set of non-toxic text, and $\barcalS$ is the set of toxic texts.
For mathematical procedures, we consider $\calS$ and $\barcalS$ to represent ``all'' non-toxic and toxic text samples.
More examples of $\calS$ and $\barcalS$ are seen in \refs{S:E}.
\end{example}

One simple idea of achieving the alignment goal is to force the text generation to stop when the generated text $x$ turns undesirable, i.e., $x\in\barcalS$, but this method involves a strong intervention in the baseline LLM, which renders the original capabilities of the baseline LLM meaningless. 
To overcome the drawback, we make the intervention strength adjustable, which enables the text-generation system to achieve the alignment goal with a \textit{weak} intervention.

\begin{figure}[h]
    \centering
    \includegraphics[width=0.8\linewidth]{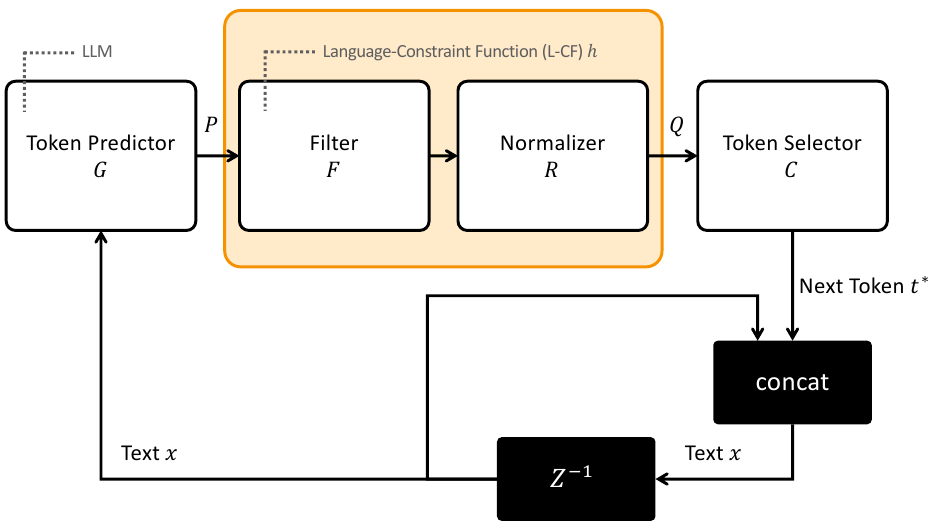}
    \caption{Structure of presented text-generation system, named CBF-LLM}
    \label{F:CBFLLM.Structure}
\end{figure}

The presented text-generation system, including an LLM and a safety filter, which is constructed based on the CBF described in \refss{SS:CBF}. 
The overall system is called CBF-LLM and its structure is shown in \reff{F:CBFLLM.Structure}.

CBF-LLM extends the nominal text-generation system shown in \reff{F:PL.NominalLLM} by adding the safety filter (yellow box) added between the token predictor $G$ and the token selector $C$.
The safety filter manipulates the probability $P$ to satisfy the specified alignment task.
The safety filter is composed of filter $F$ and normalizer $R$. In the same manner as \refss{SS:LLM}, the blocks of $G$, $C$, $\Concat$, and $Z^{-1}$ represent the token predictor, token selector, concatenator, and time delay, respectively.
The components of CBF-LLM are described in detail as follows.

Recall that the token predictor $G$ mainly implies a generative language model such as LLM.
It retrieves the current text $x\in\calX$ and outputs the probability of the next token, $P\in\bbR^N$.
For each token $t \in\calT$, the probability $P[t]$ indicates how probable each token $t$ is followed after the text $x$.

The defining feature of the CBF-LLM is the presence of filter $F$, which filters $P$ to generate the modified probability $Q\in\bbR^N$. 
The filter $F$ is designed to ensure the desirable text generation, i.e., $x\in\calS$.
To this end, we design the CBF filter in $F$ by using the function $h:\calX\to\bbR$ such satisfying
\begin{align}\begin{cases}
    h(x)\ge 0 , & x \in\calS , \\
    h(x)<0  , &  x \in\barcalS .
\label{E:CBFLLM.LCF}
\end{cases}\end{align}
The function $h$ is called the ``language-constraint function'' (L-CF).
Note that the L-CF $h$ needs to be designed to distinguish between the desired and undesired texts accurately.
We also assume that the value of L-CF $h(x)$ changes depending on the content of the text $x$.
Suppose that the alignment goal is set to generate non-toxic content. 
Then, prepare multiple samples of non-toxic text and toxic text and train a classification language model to prepare the L-CF $h$.
Specifically, the model needs to learn the relationship between non-toxic text and toxic text.
We use L-CF $h(x)$ to determine whether the text $x$ is toxic. Furthermore, the value of L-CF is expected to indicate how toxic the text is.
An example of constructing the L-CF is provided as follows.
\begin{example}
To construct L-CF, we apply a sentiment analysis RoBERTa model \footnote{\texttt{cardiffnlp/twitter\_roberta\_base\_sentiment\_latest} \cite{Loureiro22}}as the internal model.
The RoBERTa model is originally trained to classify sentences into 3 labels: negative, neutral, or positive \cite{Loureiro22}.
Let $M:\calX\to\bbR^3$ denote the RoBERTa model, and $s\in[0,1]^3$ denotes the softmax output of the $M$ respect to a text $x$, i.e., $s = \mathrm{softmax}(M(x))$.
It follows that $s[1], s[2], s[3]$ represent the score of negative, neutral, and positive, respectively.
Then, L-CF is constructed as follows:
\begin{align}
    h(x) = s[3] - \max(s[1], s[2]).
\end{align}
The function $h$ outputs a positive value when the positive score is greater than both negative and neutral scores,
while it outputs a negative value when either the negative or neutral score is greater than the positive score.
In other words, the sets $\calS$ and $\barcalS$, which correspond to the L-CF constructed above, render the positive and non-positive texts, respectively.
\end{example}

The filter $F:\bbR^N\to\bbR^N$ allows only tokens that meet its conditions to pass through and do not allow tokens that do not. 
In this paper, the CBF filter discussed in \refss{SS:CBF} is employed in $F$ and is denoted by $\FCBF$. The detailed realization of the CBF filter is given as follows:
\begin{align} 
\FCBF(P; x) : P'[t] = \begin{cases}
    P[t] , & h(\Concat(x, t)) - h(x) \le - \alpha h(x) , \\
    0 , & \text{else}
\end{cases}
,~ t \in\calT ,
\label{E:CBFLLM.CBFFilter}
\end{align}
where $\alpha$ is a hyperparameter.
This formulation is a modified form of the discrete-time CBF inequality, as shown in \eqref{E:PL.DiscreteTimeCBFConstraint}.
In \eqref{E:CBFLLM.CBFFilter}, the probability of the token is set to 0 unless the token satisfies the CBF inequality, which guarantees that the generated text $x$ always satisfies that $x\in\calS$.
\begin{remark}
    The hyperparameter in the CBF filter, $\alpha\in[0,1]$ implies the strictness of the safety constraint \eqref{E:PL.DiscreteTimeCBFConstraint}.
    In other words, the value determines the degree to which the generated text is allowed to approach the boundary of the safety constraint, $x\in\calS$.
    The CBF filter with $\alpha=1$ is the mildest: it always allows the text unless the given text is in the undesirable set, i.e., $x\in\barcalS$, 
    while the CBF filter with $\alpha=0$ is the most strict: it only allows if the text $x(k)$ is more desirable than the one in the previous time $x(k-1)$.
    In the CBF filter-based control in the assisted vehicle, the value of $\alpha$ affects the safety margin from an obstacle.
\end{remark}

To improve the computational efficiency, top-k sampling is applied in the filter $F$. The top-k sampling only processes a much smaller number of elements than $N$ elements of the target $P$.
The algorithm of the CBF filter with top-k sampling is provided in \refa{A:CBFLLM.CBFFilter}.
In the algorithm, we call that the token $t$ is \textit{allowed} (\textit{disallowed}) if the CBF inequality \eqref{E:PL.DiscreteTimeCBFConstraint} holds (does not hold) at the current text $x$.
\begin{algorithm}[h]
\caption{CBF filter $\FCBF$ with top-k sampling}
\label{A:CBFLLM.CBFFilter}

\begin{algorithmic}[1]
    \REQUIRE $P \in\bbR^N$ : token probabilities from the token predictor $G$.
    \REQUIRE $x \in\calX$ : current text.
    \REQUIRE $\alpha \in[0,1]$ : CBF's hyperparameter.
    \REQUIRE $h :\calX\to\bbR$ : the language-constraint function.
    \REQUIRE $\ktop$ : the top-k parameter.
    \STATE $P' \in\bbR^N \gets 0_N$
    \STATE $I \in\{1,\ldots,N\}^N \gets \texttt{argsort}(P)$
    \COMMENT{Sort the indexes of $P$ in descending order, i.e., $P[I[i]] \ge P[I[i+1]]$ holds for every $i\in\{1,\ldots,N-1\}$.}
    \STATE $j \gets 1$
    \STATE $k \gets 0$
    \COMMENT{Counter of allowed token}
    \WHILE{$k<\ktop$}
        \STATE $x^+ \gets \Concat(x, I[j])$
        \IF{$h(x^+)-h(x) \ge -\alpha h(x)$}
            \STATE
            \COMMENT{This token $I[j]$ is allowed: it satisfies the CBF constraint \eqref{E:PL.DiscreteTimeCBFConstraint}.}
            \STATE $P'[I[j]] \gets P[I[j]]$
            \STATE $k \gets k + 1$
        \ENDIF
        \STATE $j \gets j + 1$
    \ENDWHILE
    \RETURN $P'$
\end{algorithmic}
\end{algorithm}

The normalizer $R$ adjusts the output of the filter $P'$ to ensure that it is normalized, such that the sum of the output $Q$ equals 1. 
\begin{align}
    R : Q[i] = \frac{P'[i]}{\sum_{j=1}^N P'[j]}
    ,\quad i\InRange{1}{N}
\end{align}

The CBF-LLM given in \reff{F:CBFLLM.Structure} with the identity filter $F(P)=P$ instead of the CBF filter $\FCBF$ is reduced the nominal text-generation system, given in \reff{F:PL.NominalLLM}.
The algorithm of the nominal text-generation system with top-k sampling is stated in \refappendix{S:TopKSamplingNominalTextGeneration}, and it is implemented in \refs{S:E} for comparison with CBF-LLM.

\begin{remark}
    We emphasize that the CBF-LLM, the text-generation system presented in \reff{F:CBFLLM.Structure}, provides a framework for aligning LLMs in a different manner from traditional approaches \cite{Ouyang22}\cite{Shen23}\cite{Dong24}.
    In traditional approaches, training datasets are usually utilized to train 
    reward models in RLHF \cite{Ouyang22},
    LLMs in SFT \cite{Shen23}, or
    directly put on the prompt in the in-context learning \cite{Dong24}.
    The datasets tell what the desirable texts and undesirable texts are.
    In CBF-LLM, the role of the datasets is taken over by the L-CF, as shown in \eqref{E:CBFLLM.LCF}. 
    The L-CF is assumed to accurately classify the text into desirable $\calS$ or undesirable $\barcalS$, based on the respective alignment task purpose.
\end{remark}




\section{Experiment}\label{S:E}
To demonstrate the CBF-LLM's alignment ability and the intervention time, we implement CBF-LLM by pre-trained large language models.

\subsection{Setting}\label{SS:E.Setting}
In the experiment, the alignment goal is to ensure that the text-generation system, illustrated as in \reff{F:CBFLLM.Structure}, does not produce any ``non-positive'' output.
To this end, we let $\calS$ and $\barcalS$ denote the set of positive texts and non-positive texts, respectively.
We employ Llama~3~8b \cite{Dubey24} as the model for the token predictor $G$, 
and \texttt{cardiffnlp/twitter\_roberta\_base\_sentiment\_latest} \cite{Loureiro22}, a sentiment analysis RoBERTa model, as the language-constraint function (L-CF) $h$.
The internal model of the L-CF was originally trained to classify sentences into 3 labels: negative, neutral, or positive \cite{Loureiro22}.
In this experiment, we apply the following mapping to construct the L-CF:
\begin{align}\begin{cases}
    s = \mathrm{softmax}(M(x)) ,\\
    h(x) = s[3] - \max(s[1], s[2]),
\end{cases}\end{align}
where $M:\calX\to\bbR^3$ is the internal model 
and $s[1]$, $s[2]$, and $s[3]$ represent the score of negative, neutral, and positive, respectively.
The function $h$ outputs a positive value when the sentiment of the text $x$ is positive.
It means that the text-generation system would be controlled to generate positive content.

In the text generation experiment with CBF-LLM, we applied each of the following filters as $F$.
\begin{description}
    \item[CBF($\alpha=0.3$)] Filter with the CBF with hyperparameter $\alpha=0.3$, which implies that the executable area is small, in other words, the CBF constraint \eqref{E:PL.DiscreteTimeCBFConstraint} is strict. The algorithm is shown in \refa{A:CBFLLM.CBFFilter}.
    
    \item[CBF($\alpha=0.8$)] Filter with the control barrier function with hyperparameter $\alpha=0.8$, which implies that the executable area is wider, in other words, the CBF constraint \eqref{E:PL.DiscreteTimeCBFConstraint} is more moderate. The algorithm is shown in \refa{A:CBFLLM.CBFFilter}.
    
    \item[Blacklist] 
    The Blacklist filter disallows tokens such that the L-CF $h$ for concatenated text $\Concat(x,i)$ indicates a negative value.
    The Blacklist filter is equivalent to the CBF filter with the hyperparameter $\alpha=1$.
    
    \item[NoControl] No filtering is applied to probabilities $P$ is performed, i.e., $F$ is the identity map and the proposed text-generation system as shown in \reff{F:CBFLLM.Structure} is reduced to the traditional one, as shown in \reff{F:PL.NominalLLM}. 
    The system is expected to be operated only by the baseline model, Llama~3.
    The algorithm is shown in \refappendix{S:TopKSamplingNominalTextGeneration}.
\end{description}

In this experiment, we set the 
initial text as ``Everyone says you will be a good researcher in the future, but'',
temperature as $T=1$,
top-k value as $k_\mathrm{top}=30$,
and the maximum number of new tokens as 30.
We generated 100 samples.


\newcommand\GeneratedTextListWContainer[1]{%
\begin{minipage}{0.9\textwidth}\begin{itemize}%
#1%
\end{itemize}\end{minipage}}

\subsection{Result}
The following are examples of generated texts by each filter. The red slash \LineBreak indicates the line break.

\noindent CBF($\alpha=0.3$)
\begin{itemize}
\item Everyone says you will be a good researcher in the future, but you are already such a good one now.\LineBreak It was my pleasure to get to know you when I came to the University for my fieldwork. I 
\item Everyone says you will be a good researcher in the future, but you should also read more books, to be an excellent researcher.” I have heard these words repeatedly from senior researchers at many academic events and seminars. Although 
\item Everyone says you will be a good researcher in the future, but you do know the secret of research, do you. A researcher is someone who is curious, and a curious person is an interested one.\LineBreak If you 
\item Everyone says you will be a good researcher in the future, but I have to say you are so cute and charming as a lecturer. So, I thought what you say is very important to our understanding of a new 
\end{itemize}

\noindent CBF($\alpha=0.8$)
\begin{itemize}
\item Everyone says you will be a good researcher in the future, but I want to encourage you to be a good researcher now,” he said. “I remember reading somewhere that ‘good research is always useful.’ You can 
\item Everyone says you will be a good researcher in the future, but there's a catch with this. You'll have an extremely huge field of study. It's good, and it also opens doors for you.\LineBreak I 
\item Everyone says you will be a good researcher in the future, but I also hope that you will find happiness. That is also very important. But I know that you cannot be happy when you are busy working all day 
\item Everyone says you will be a good researcher in the future, but you can learn a lot if you take part in teaching and administration.\LineBreak I enjoy being able to meet, interact with and teach the next generation of researchers 
\end{itemize}

\noindent Blacklist 
\begin{itemize}
\item Everyone says you will be a good researcher in the future, but I donot belie... \LineBreak You can write a book, so can everyone. \LineBreak There is always plenty to learn, I want my life... \LineBreak I am happy 
\item Everyone says you will be a good researcher in the future, but in fact, your work is only a small piece of a big piece of work.\LineBreak The world is too great and your work is too tiny. 
\item Everyone says you will be a good researcher in the future, but we're very different. I am the type of person who can go outside and find something to study," Kelli added. "You, on your 
\item Everyone says you will be a good researcher in the future, but my question is - what is your goal?\LineBreak This is what you should know about a person, his future career ambition and development and in how many ways 
\end{itemize}

\noindent NoControl 
\begin{itemize}
\item Everyone says you will be a good researcher in the future, but you are not so much. Why?\LineBreak \LineBreak What advice could you give to others who are studying the same subjects as you?\LineBreak \LineBreak What has been your biggest failure
\item Everyone says you will be a good researcher in the future, but we think it is not enough, you must become an authority in the field. This is not just to say that you will become a research, but 
\item Everyone says you will be a good researcher in the future, but I do not believe in the future," said one young researcher in a conversation I had in Delhi in 2005. This young and disillusioned researcher 
\item Everyone says you will be a good researcher in the future, but right now, your research is still at the infant's stage. I am sure you will be able to do better if you are working on a topic 
\end{itemize}

In the NoControl case, where no alignment is applied in the text-generation system, some texts are going non-positive content. 
This implies that the baseline LLM, Llama 3, can generate non-positive texts.
With the Blacklist filter and the CBF filter, all texts are positive content.

\begin{figure}[h]
    \centering
    \includegraphics[width=1\linewidth, clip]{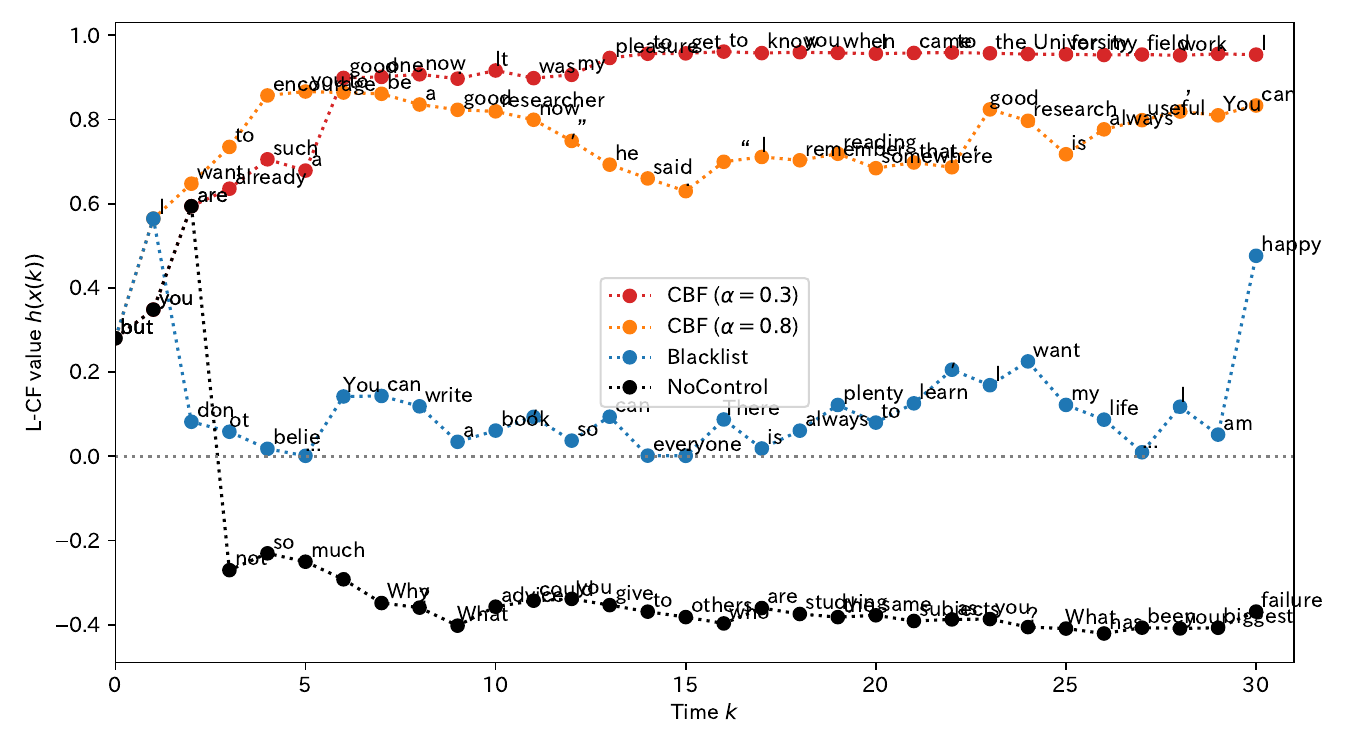}
    \caption{L-CF trajectory of each controller}
    \label{F:LCFTrajectories}
\end{figure}

\reff{F:LCFTrajectories} shows the trajectory of language-constraint function $h(x(k)), k\in\{1,2,\ldots\}$.
In the NoControl case (black line), the generated text does not keep the positive L-CF value, implying the extent to which the generated text is undesirable.
On the other hand, CBF filters (red line and orange line), the positive L-CF values are kept during the generation, 
implying that the text generation system generates desirable content.
The Blacklist filter (blue line) also maintains a positive L-CF value, but it frequently selects tokens near $h=0$.

\begin{figure}[h]
    \centering
    \includegraphics[width=1\linewidth]{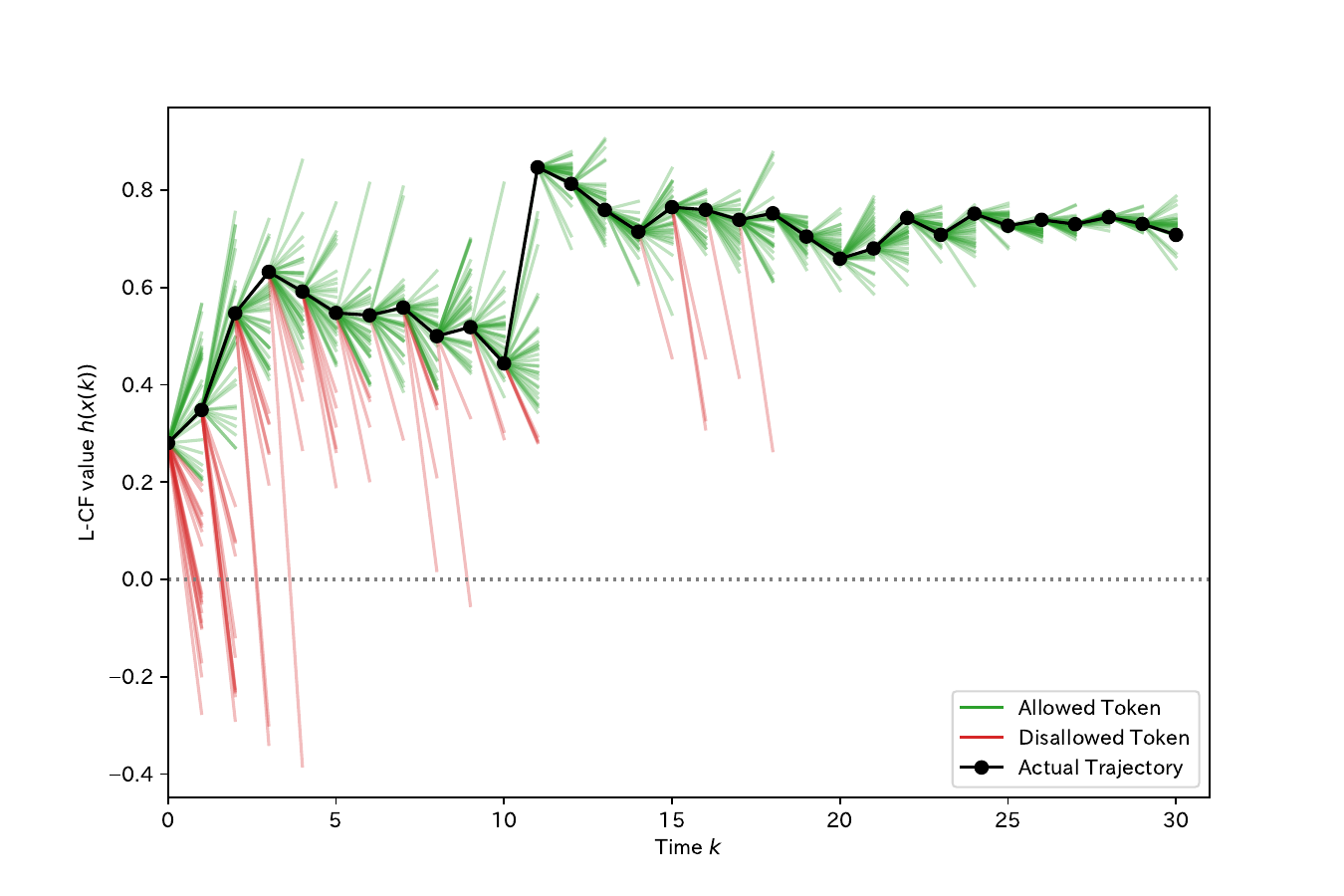}
    \caption{Predicted L-CF trajectories}
    \label{F:LCFPredTrajectories}
\end{figure}

\reff{F:LCFPredTrajectories} shows the predicted ``possible'' trajectoroes of the CBF filter $\FCBF(\alpha=0.3)$. 
In this example, the resulting text was ``Everyone says you will be a good researcher in the future, but you should also read more books, to be an excellent researcher.” I have heard these words repeatedly from senior researchers at many academic events and seminars. Although''.
Recalling \refa{A:CBFLLM.CBFFilter}, at each time $k$, the CBF filter sorts tokens into those that satisfy the CBF inequality and those that do not. 
It is shown that the CBF filter prevents L-CF values from becoming negative or decreasing more rapidly than the current value.
Note that we do not show the trajectories for all tokens, but only for tokens investigated by the top-k sampling (see \refa{A:CBFLLM.CBFFilter}) are displayed.

Next, we discuss the number of interventions. 
The CBF-LLM aims to avoid frequent intervention in the baseline LLM output.
\reft{T:E.DiallowedTokens} shows the average number of disallowed tokens in a single generation. 
It is clear through this experiment that the disallowed tokens in the CBF filter are less than that in the blacklist filter. 
This implies that the CBF filter would contribute to the reduced intervention.
\begin{table}[h]
    \centering
    \caption{Disallowed tokens}
    \label{T:E.DiallowedTokens}
    \begin{tabular}{|c|c|c|c|c|}
        Filter & NoControl & Blacklist & CBF($\alpha=0.8$) & CBF($\alpha=0.3$) \\
    \hline
        \# of disallowed tokens & 0 & 209.79 & 137.90 & 161.59 \\
    \end{tabular}
\end{table}

\begin{figure}[h]
    \centering
    \begin{subfigure}{0.42\textwidth}
        \centering
        \includegraphics[width=\textwidth]{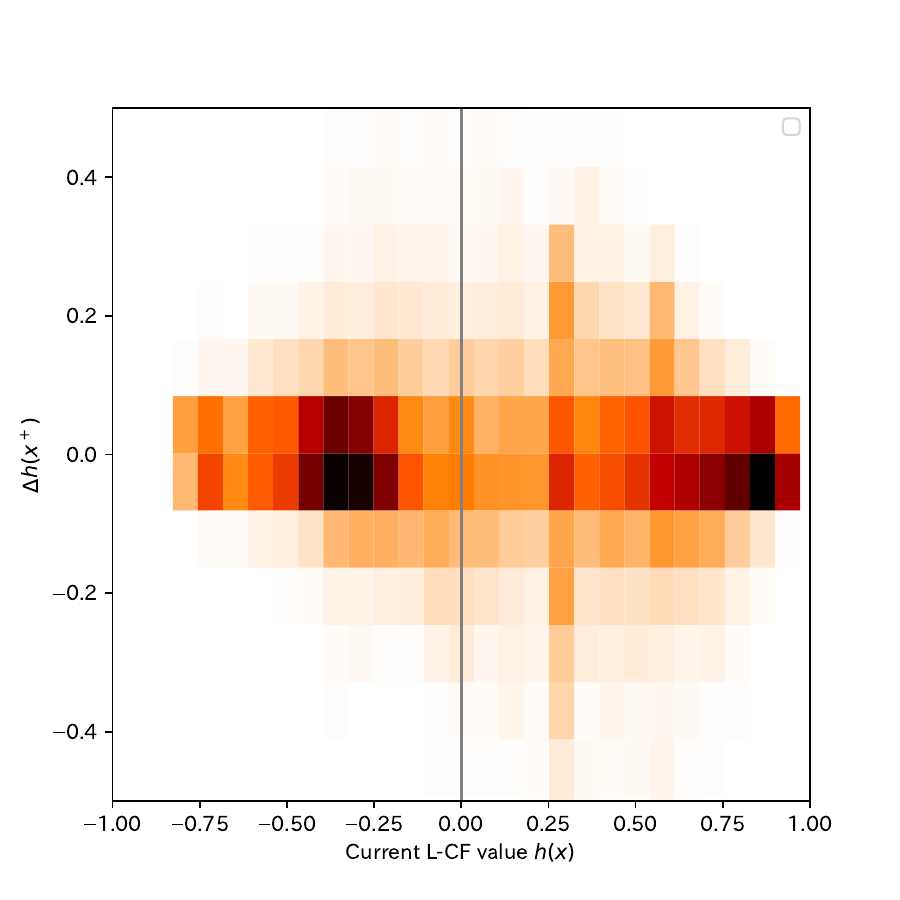}
        \caption{NoControl}
        \label{F:E.Attractors.NC}
    \end{subfigure}
    \begin{subfigure}{0.42\textwidth}
        \centering
        \includegraphics[width=\textwidth]{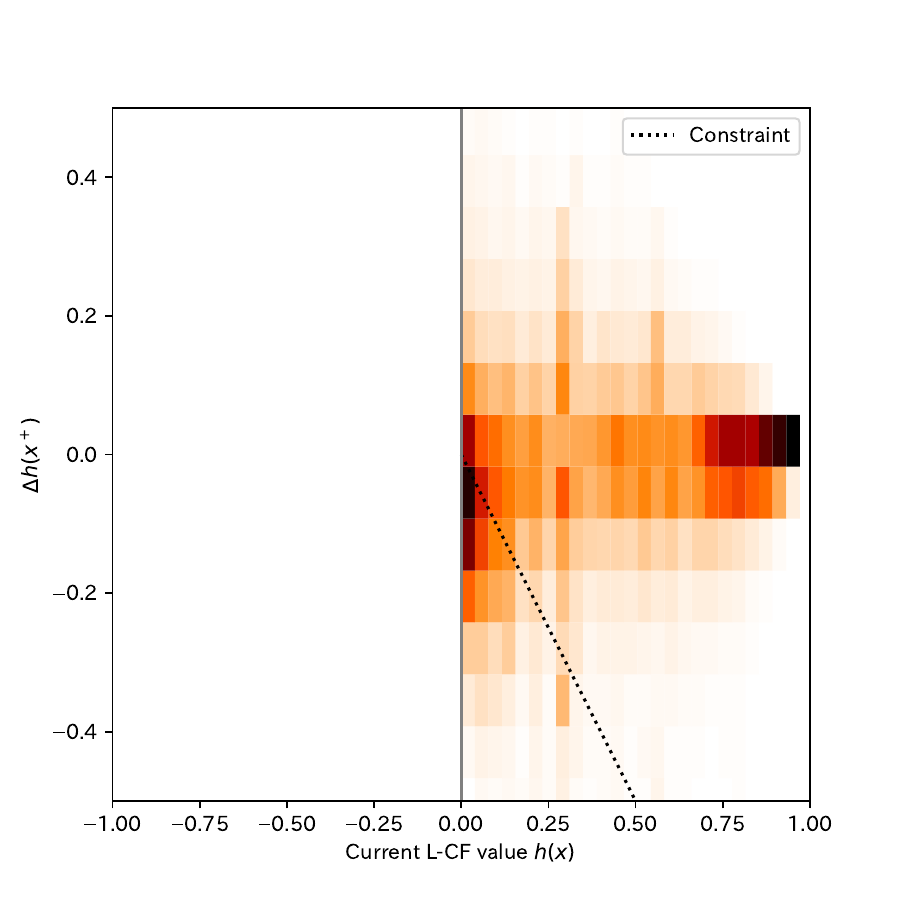}
        \caption{Blacklist}
        \label{F:E.Attractors.BL}
    \end{subfigure}
    \begin{subfigure}{0.42\textwidth}
        \centering
        \includegraphics[width=\textwidth]{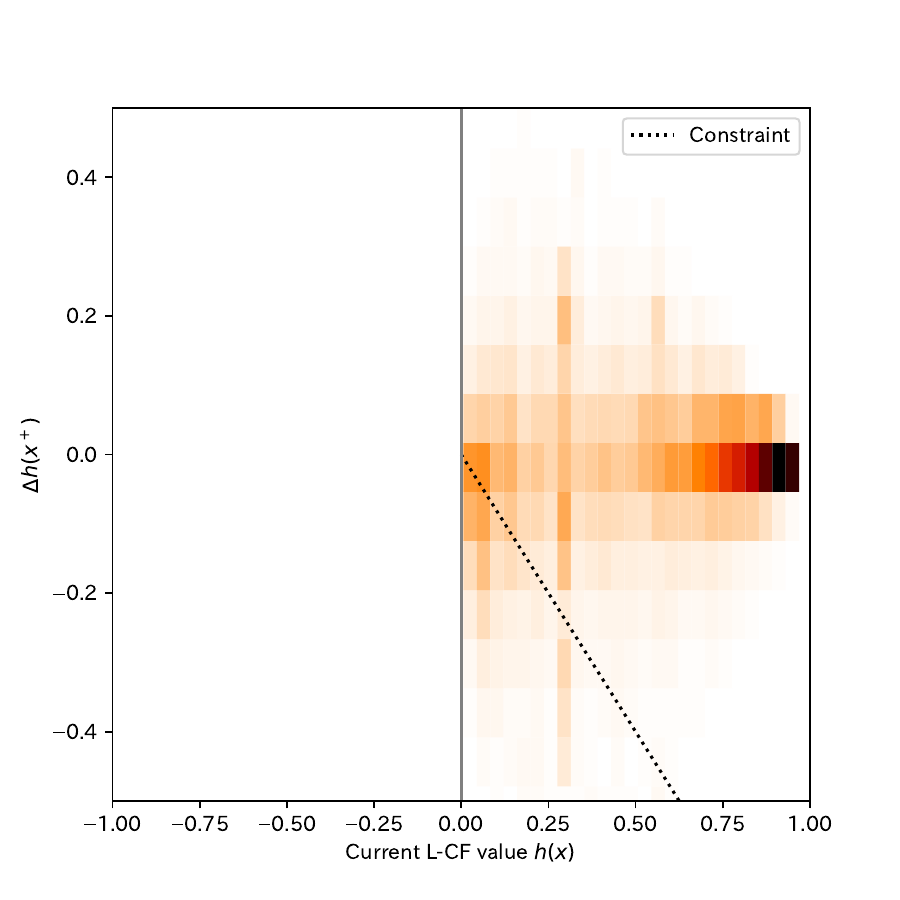}
        \caption{CBF($\alpha=0.8$)}
        \label{F:E.Attractors.CBF0.8}
    \end{subfigure}
    \begin{subfigure}{0.42\textwidth}
        \centering
        \includegraphics[width=\textwidth]{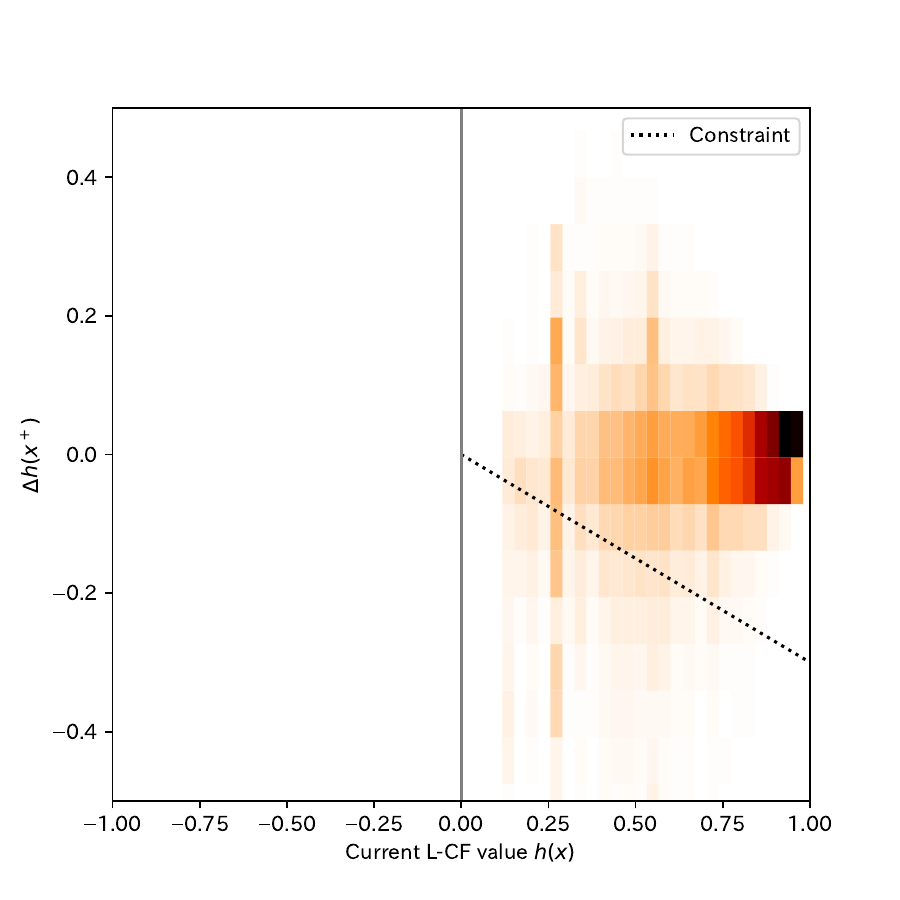}
        \caption{CBF($\alpha=0.3$)}
        \label{F:E.Attractors.CBF0.3}
    \end{subfigure}
    
    \label{F:E.Attractors}
    \caption{Attractors: 2D histogram of $(h,\Delta h)$}
\end{figure}

In \reff{F:LCFTrajectories}, the values of $h$ tend to remain within a close range.  
This resembles the formation of ``attractors'', the concept of dynamical systems \cite{Guckenheimer83}. 
To verify this observation, we conduct a supplemental experiment of studying the distribution of the value of $h$ with respect to each filter.
Figs.~\ref{F:E.Attractors.NC}, \ref{F:E.Attractors.BL}, \ref{F:E.Attractors.CBF0.8}, and \ref{F:E.Attractors.CBF0.3} show the distribution of the L-CF value $h$ while the generation of each filter.
In the figures, the horizontal axis is the value of $h$, and the vertical axis is the difference in $h$ from the previous time step.
The figures show the results of analyzing all tokens processed by the filter during the generation process.
Their plots are the value of $h$ and its difference for each token when it is selected.
The dotted lines represent the CBF constraint, and tokens with $\Delta h$ value lower than them are disallowed by the filters.
The values of $h$ tend to remain within a close range. 
In the NoControl and the Blacklist cases, the values of $h$ tend to cluster around two distinct attractors.
However, in the CBF with $\alpha=0.8$ case, the values of $h$ tend to cluster around strong mode and shallow attractor, and
in the CBF with $\alpha=0.3$ case, the values of $h$ tend to cluster around a single attractor.
In \reff{F:E.Attractors.NC}, some tokens cluster in the negative range of $h$, implying that the baseline LLM tend to generate the non-positive content.
These results imply that the attractor of the L-CF value $h$ gets influenced by the CBF filter and its hyperparameter value, $\alpha$.



\section{Conclusion}\label{S:Conclusion}
This paper proposed the control-based LLM alignment framework, called CBF-LLM.
This framework utilizes the control barrier function (CBF), commonly used in control engineering to ensure the safety of physical objects, such as the collision avoidance function in assisted driving vehicles.
Based on an analogy between the control theory and the LLM alignment task,
We employ the CBF-based safety filter to ensure that the text-generation system generates desirable content.
The key feature of CBF-LLM is that the CBF filter is an add-on to the baseline LLM: 
it intervenes in the output of the baseline LLM without any additional training of LLMs.

This paper also presented the implementation of CBF-LLM by Llama~3 and a sentiment analysis RoBERTa model to ensure that the text-generation system generates positive content.
The text-generation experiment showed that CBF-LLM successfully achieves the alignment goal with fewer interventions than the blacklist method.

The future works include conducting text-generation experiments with other LLMs and other alignment goals,
and theoretical insights on the number of interventions by the CBF filter of CBF-LLM.

\bibliographystyle{ieeetr}
\bibliography{References}


\newpage
\appendix
\noindent {\large \textbf{Appendix}}

\section{Nominal Text Generation with Top-k Sampling}\label{S:TopKSamplingNominalTextGeneration}
Recall that \reff{F:CBFLLM.Structure} shows the text-generation system and intervention procedure in block diagram format.
To represent general top-k sampling with no control in this figure, the filter $\FNC$ is provided, as shown in \refa{A:Appendix.TopKSamplingNominalTextGeneration}.
\begin{algorithm}[h]
\caption{No-control filter $\FNC$ with top-k sampling}
\label{A:Appendix.TopKSamplingNominalTextGeneration}
\begin{algorithmic}[1]
    \REQUIRE $P \in\bbR^N$ : token probabilities from the token predictor $G$.
    \REQUIRE $x \in\calX$ : current text.
    \REQUIRE $\ktop$ : the top-k parameter.
    \STATE $P' \in\bbR^N \gets 0_N$
    \STATE $I \in\{1,\ldots,N\}^N \gets \texttt{argsort}(P)$
    \STATE $k \gets 1$
    \WHILE{$k<\ktop$}
        \STATE $P'[I[k]] \gets P[I[k]]$
        \STATE $k \gets k + 1$
    \ENDWHILE
    \RETURN $P'$
\end{algorithmic}
\end{algorithm}


\end{document}